\documentclass[a4paper]{article}
\usepackage{wds10,epsf}

\usepackage[square]{natbib}
\usepackage{graphicx}
\usepackage{color}
\usepackage{enumitem}

\lefthead{GEROLD BUSCH}
\righthead{Black hole - host relations in active galaxies}

\begin{document}

\title{Correlations between Supermassive Black Holes and their Hosts in Active Galaxies}

\author{Gerold Busch}

\affil{I. Physikalisches Institut der Universit\"at zu K\"oln, Cologne, Germany}

\begin{abstract}
In the last decades several correlations between the mass of the central supermassive black hole (BH) and properties of the host galaxy - such as bulge luminosity and mass, central stellar velocity dispersion, S\'ersic index, spiral pitch angle etc. - have been found and point at a coevolution scenario of BH and host galaxy. In this article, I review some of these relations for inactive galaxies and discuss the findings for galaxies that host an active galactic nucleus/quasar. I present the results of our group that finds that active galaxies at $z\lesssim 0.1$ do not follow the BH mass - bulge luminosity relation. Furthermore, I show near-infrared integral-field spectroscopic data that suggest that young stellar populations cause the bulge overluminosity and indicate that the host galaxy growth started first. Finally, I discuss implications for the BH-host coevolution. 
\end{abstract}

\begin{article}

\section{Introduction}

Black holes (BH) are amongst the most extreme objects in the Universe. On the other hand, they are by definition not directly observable which makes them one of the most fascinating objects of study. While they have been found as solutions of the theory of General Relativity already in the beginning of the 20th century, the name ``black hole'' came up only in the 60s. At the same time the idea that supermassive (i.e. millions to billions times more massive than our Sun) black holes reside in the centres of galaxies and are responsible for the powering of quasars became popular\footnote{For a detailed review on the historical development from purely mathematical ideas to first observations of black holes, including many references, the reader is referred to the articles by \cite{1987thyg.book..199I} and \cite{2016ASSL..418..263G}, chapters 2 and 3.}.

However, the ultimate proof that the compact, dark masses that observations suggest in the centres of many galaxies were black holes is still outstanding. Only in the beginning of this century, the intense effort spent on observation campaigns of stellar orbits around the central massive object of our Milky Way at highest angular resolution, carried out over more than ten years, paid back: Due to the high density of the central object with a mass of a few million solar masses, many alternatives such as dark clusters of low-mass stars, neutron stars or stellar black holes can be ruled out, strongly supporting the existence of a supermassive black hole the first time \citep[][but see also the upcoming paper ``The Galactic Centre Black Hole - A challenge for Astronomy and Philosophy'' by A.~Eckart et al.]{2002MNRAS.331..917E,2002Natur.419..694S}. Due to its proximity, the Galactic Centre is still the ultimate laboratory for testing physics in the vicinity of a Supermassive Black Hole \citep[e.g.][see also talks by Eckart, Karssen, Parsa, Zaja\v{c}ek at this conference]{2015arXiv150102171E}.

By now, black hole masses have been measured directly from the stellar dynamics, the ionised gas or maser dynamics, or reverberation mapping (for unobscured AGN) for around 80 galaxies\footnote{For a review on the historical development of black hole mass measurements, the reader is referred to the article by \cite{2013ARA&A..51..511K}, chapters 2 and 3.}. Already in the beginning of these measurement campaigns, \cite{1988ApJ...324..701D} compared the $M_\mathrm{bulge}/M_\mathrm{BH}$ and $M_\mathrm{total}/M_\mathrm{BH}$ ratios of two objects and first mentioned that the BH mass may be correlated with the bulge luminosity rather than the total host galaxy luminosity. Further objects were added in the following years, the review by \cite{1995ARA&A..33..581K} already contained eight objects. The first systematic search, resulting in an unbiased sample of 32 galaxies, was undertaken by \cite{1998AJ....115.2285M}, resulting in a single power-law $M_\mathrm{BH}-M_\mathrm{bulge}$ relation. With more and better measurements since then, the scatter of the relation has been reduced significantly and several other relations between the BH and its host galaxy have been found. Because of the different spatial and mass scales involved, these relations are not trivially expected. Instead, they suggest an intimate coupling of the evolution of black holes and their host galaxies (or at least their central spheroids/bulges). An open question is how this coevolution works: Is it the feedback of the black hole that regulates the growth of the surrounding host galaxy or is it the stellar and gaseous environment that regulates the growth of black holes? Understanding the location and evolution of galaxies that harbour an active galactic nucleus, i.e. whose central BH is in a growing phase, in BH - host galaxy relations can help us solving this question.

BH mass - host galaxy scaling relations are an important tool to determine BH masses for larger samples and study the evolution of BHs, especially in the context of the evolution of their host galaxies. Furthermore, scaling relations give important constrains for simulations of galaxy evolution.

In this article, I first review several known BH - host galaxy relations for inactive galaxies and also discuss their difficulties. Then, I present our results for active galaxies that deviate from these relations and explain implications on the BH - host galaxy coevolution.

\section{$M_\mathrm{BH}-L_\mathrm{bulge}$ and $M_\mathrm{BH}-M_\mathrm{bulge}$ relation}

An important step to lower the scatter in the $M_\mathrm{BH}-L_\mathrm{bulge}$ relation was going to near-infrared wavelengths which better trace the mass-dominating old stellar populations and are less affected by dust obscuration \citep{2003ApJ...589L..21M}. The first studies were mostly based on massive early-type galaxies\footnote{Elliptical and lenticular galaxies are commonly called early-type galaxies, while spiral and irregular galaxies are called late-type. The names come from the original interpretation of the ``Hubble tuning fork'' that assumes an evolution from elliptical galaxies to spiral galaxies.} where the spheroid luminosity can be determined in a rather simple way. Late-type galaxies, however, consist of several components like disc, bulge, and sometimes a bar. This requires a morphological decomposition where usually analytic functions are used to model the surface brightness. A general expression that can be used to describe the profiles is the S\'ersic function \citep{1968adga.book.....S,1993MNRAS.265.1013C}:
\begin{equation}
\mu_\mathrm{Sersic}(r;\mu_e,r_e,n) = \mu_e + \frac{2.5\,b_n}{\ln(10)} \left[ \left( \frac{r}{r_e} \right)^{1/n} -1 \right]
\end{equation}
where $\mu_e$ is the surface brightness at the effective radius $r_e$, the S\'ersic index $n$ measures the concentration of light, and $b_n$ is a function of $n$ chosen such that half of the luminosity is contained within $r_e$ (``half-light radius''). For $n=1$ the function reduces to an exponential profile (which can be used to fit the disc component), for $n=1/2$ to a Gaussian function (bar component) and for $n=4$ to a de Vaucouleurs profile which is the ``historic'' light profile for elliptical galaxies and bulges. Nowadays, bulges are usually fitted with a free S\'ersic parameter $n\approx4$. For the decomposition, several two-dimensional codes are available, the most used ones are galfit \citep{2010AJ....139.2097P}, BUDDA \citep{2008MNRAS.384..420G}, and IMFIT \citep{2015ApJ...799..226E}. However, some authors also perform one-dimensional fits \citep[recently][]{2016ApJS..222...10S}. One has to note that the aforementioned decompositions are very sensitive to factors like the background subtraction and (especially in the case of AGN with a strong point source) a good knowledge of the point-spread-function. These effects can be modelled and the uncertainties thereby estimated \citep[e.g.][]{2014A&A...561A.140B}. More difficulties are caused by the fact that it is not known a-priori which morphological components have to be included in the fit (spatially resolved spectroscopy can be a help here) and in some cases different models yield equally good fits. The uncertainty introduced by this degeneracy is difficult to estimate.

Most authors find that the BH mass correlates better with the bulge luminosity than with the total host galaxy luminosity (except for early-type galaxies) with $M_\mathrm{BH} \propto L_\mathrm{bulge}$ \citep[e.g.][]{2012MNRAS.419.2264V,2013ApJ...764..184M,2013ARA&A..51..511K}. One recent exception is \cite{2014ApJ...780...70L} who find that bulge luminosity and total host luminosity correlate equally well with the BH mass, probably due to their decomposition that includes a multitude of additional components (what makes it difficult to define what the ``bulge'' is) and their large fraction of early-type galaxies \citep[only 4 out of 35 objects are late-type galaxies; see][]{2016ApJ...817...21S}.

A correlation between BH mass and stellar (bulge) mass is in particular important for comparison with simulations. \cite{2003ApJ...589L..21M} show that the virial mass $M_\mathrm{bulge} = k r_e \sigma^2$ follows a similar linear correlation with BH mass as the NIR luminosity. Improved dynamical measurements have been delivered by \cite{2004ApJ...604L..89H}. \cite{2013ARA&A..51..511K} point out that virial estimates are problematic due to the uncertainties in the determination of $r_e$ and the choice of the geometrical factor $k$. The other possibility is to derive bulge masses from the bulge luminosity by applying an appropriate mass-to-light ratio, $M/L$. However, even in the NIR mass-to-light ratios are not easy to obtain, for example due to the unknown IMF and possible age-metallicity degeneracies. In particular, young stellar populations expected in active galaxies and AGB stars can heavily affect $M/L$ ratios but are difficult to detect \citep[e.g.][]{2014A&A...561A.140B}. Spatially resolved spectroscopy is necessary to constrain stellar population ages from which $M/L$ ratios can be derived.

\section{$M_\mathrm{BH}-\sigma_*$ relation}

A tight relation between the BH mass and the host galaxy's stellar velocity dispersion $\sigma_*$ was independently discovered by \cite{2000ApJ...539L...9F} and \cite{2000ApJ...539L..13G}. Depending on the used fitting algorithm\footnote{This effect is discussed extensively in Chapter 11.5.1 of \cite{2016ASSL..418..263G}} and the data set, authors find relations $M_\mathrm{BH} \propto \sigma_*^{3.5-6}$. While the relations were first suspected to have zero intrinsic scatter, more recent studies show that the intrinsic scatter is comparable to that of the $M_\mathrm{BH}-L_\mathrm{bulge}$ relation \citep[e.g.][]{2009ApJ...698..198G}. Several difficulties arise, especially when compiling a large sample of $\sigma_*$ measurements from the literature, some are: (i) In which aperture is $\sigma_*$ measured? Often the effective radius $r_e$ is used, but also fractions of it. (ii) Is rotation included in or subtracted from the measurement of the velocity dispersion? (iii) To what extent are substructures and additional components such as nuclear discs, pseudobulges or bars accounted for? (iv) Do pseudobulges correlate with the BH mass? \citep{2011Natur.469..374K}

\section{Pseudobulges or a bent relation?}

Of particular interest is the behaviour of low-mass galaxies, a region that is not sampled well in most of the early studies. The central components of many low-mass galaxies are so-called ``pseudobulges'', different from classical bulges by being more disky, less concentrated, with higher rotational support and containing younger stellar populations \citep{2004ARA&A..42..603K}. The classification, in particular if only based on one of these criteria, is not free of ambiguities. For example, although a trend towards lower S\'ersic indices is present, the S\'ersic index alone is not a sufficient criterion \citep{2009MNRAS.393.1531G}. In addition, pseudobulges and classical bulges can coexist in one galaxy \citep{2015MNRAS.446.4039E}.

\cite{2011Natur.469..374K} find that pseudobulges do not follow BH mass - host galaxy scaling relations. They attribute this to different BH feeding scenarios: While bulges originate from global feeding mechanisms, probably driven by dry mergers, pseudobulges originate from local, rather stochastical, mass assembly inside the disc that do not result in a coevolution of BH and host.

Alternatively, \cite{2012ApJ...746..113G} suggest to break the $M_\mathrm{BH}-L_\mathrm{bulge}$ and $M_\mathrm{BH}-M_{*,\mathrm{bulge}}$ relations. In this scheme, only luminous core-galaxies follow a linear BH - bulge relation that originates from gas-poor, ``dry'' mergers. On the other hand, S\'ersic galaxies at the low-mass end follow a near-quadratic relation that originates from gas-rich, ``wet'' processes, in which black holes grow faster than their host galaxies \citep{2013ApJ...764..151G}. A bend in the relation is expected if one assumes a linear $M_\mathrm{BH}-\sigma_*$ relation, given the fact that the $\sigma-M_{*}$ relation is broken \citep[e.g.][]{2016arXiv160204267C}.

\section{Further relations}
In addition to the presented ones, several other correlations between BH mass and host galaxy properties have been found that I cannot all review in detail. Here, two relations are shortly mentioned that have the advantage that they are based on parameters that can be derived from \emph{uncalibrated} images. In particular this means that they are independent of the assumed cosmology and distance to the object.

\paragraph{S\'ersic index $n$}
\cite{2001ApJ...563L..11G} suggested that it could not only be the mass or luminosity of the central spheroid that is related to the BH mass but also the way that the mass is distributed. \cite{2007ApJ...655...77G} present a log-quadratic relation between the central BH mass and the S\'ersic index $n$ as a measure of the stellar mass concentration, the $M_\mathrm{BH}-n$ relation. While several authors in the following could not recover a strong $M_\mathrm{BH}-n$ correlation from their multicomponent decompositions, \cite{2013MNRAS.434..387S} show that this could be due to inaccurate decompositions. In \cite{2016ApJ...821...88S}, she finds a close relation $M_\mathrm{BH} \propto n^{3.39\pm 0.15}$ between BH mass and S\'ersic index. I note that for active galaxies with strong central point source contributions, it is difficult to recover the S\'ersic index $n$ (especially in late-type galaxies with small spheroidal component). In general, since the parameters in the S\'ersic function are not independent of each other, one has to be very cautious while fitting and very high signal-to-noise data is required.

\paragraph{Spiral arm pitch angle}
Furthermore, a relation between BH mass and the pitch angle of the spiral arms has been recovered \citep[e.g.][]{2008ApJ...678L..93S,2013ApJ...769..132B}, with large pitch angles (widely opened spiral arms) correlating with smaller BHs. This relation is somehow expected since the pitch angle increases towards later Hubble types while at the same time the bulge fraction decreases which is by the $M_\mathrm{BH}-L_\mathrm{bulge}$ and $M_\mathrm{BH}-M_{\mathrm{bulge},*}$ relations correlated with the BH mass. This relation could be particularly of interest for bulgeless galaxies.

\section{BH-host galaxy relations for active galaxies}
While $M_\mathrm{BH}$-host galaxy relations are clearly very useful tools to estimate the black hole masses for large samples of galaxies, they can also help understanding galaxy evolution. Active galaxies are in a current phase of BH growth and can therefore help to search the mechanisms that drive galaxies towards (or away from) the scaling relations. 

Several authors find that active galaxies do not follow the $M_\mathrm{BH}-L_\mathrm{bulge}$ relation. While \cite{2004ApJ...615..652N} associate the deviation of about one magnitude in bulge brightness with an overluminosity of the bulge due to young stellar populations, \cite{2008ApJ...687..767K} argue that the deviation in their study is produced by still growing undermassive black holes. They further study how the magnitude of the offset depends on host and AGN properties and find that at given bulge luminosity objects with higher Eddington ratios have lower BH masses. 
\cite{2012ApJ...757..125U} study young red quasars and come to a similar conclusion: Objects with high Eddington ratios are shifted below the $M_\mathrm{BH}-L_\mathrm{bulge}$ relation of inactive galaxies, while those with low accretion rates are close or even slightly above the relation. The most favourable scenario to explain this shift from the $M_\mathrm{BH}-L_\mathrm{bulge}$ is that the host galaxy grows first, followed by BH growth some time later.

\cite{2011ApJ...737L..45J} investigate a sample of active galaxies with low-mass black holes ($10^5-10^6\,M_\odot$) and show that they fall below the linear $M_\mathrm{BH}-L_\mathrm{bulge}$ relation of inactive galaxies with an offset in bulge luminosity of $\sim 1\,\mathrm{mag}$. Their interpretation is that most of the host galaxies contain pseudobulges and are therefore not expected to follow the linear relation. However, by choice the sample falls into the low-mass regime and could therefore also be interpreted as representative of the near-quadratic $M_\mathrm{BH}-L_\mathrm{bulge}$ relation \citep{2016ASSL..418..263G}. 

Narrow-line Seyfert 1s constitute an extreme case in this context since most of them have very small host spheroidals, often with characteristics of pseudobulges \citep{2011MNRAS.417.2721O}, and have extremely high accretion rates. They deviate significantly from the $M_\mathrm{BH}-L_\mathrm{bulge}$ relation \citep{2011nlsg.confE..42W,2012ApJ...754..146M} which could imply that they are in an early evolutionary state. However, a bias in the selection of the sample cannot be discarded \citep{2012nsgq.confE..17V}.

\begin{figure}[t]
\begin{center}
\includegraphics[width=0.49\linewidth]{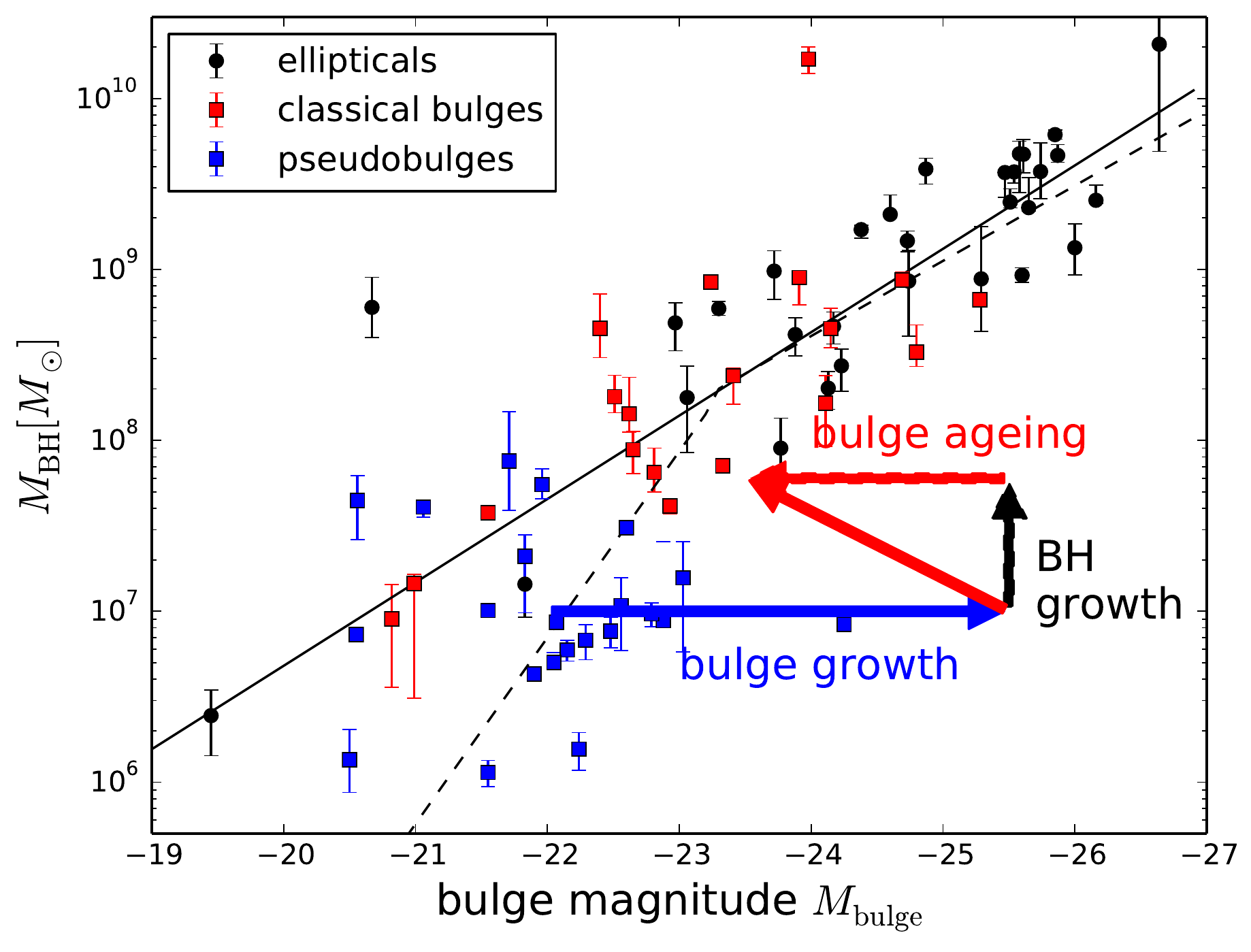}
\includegraphics[width=0.49\linewidth]{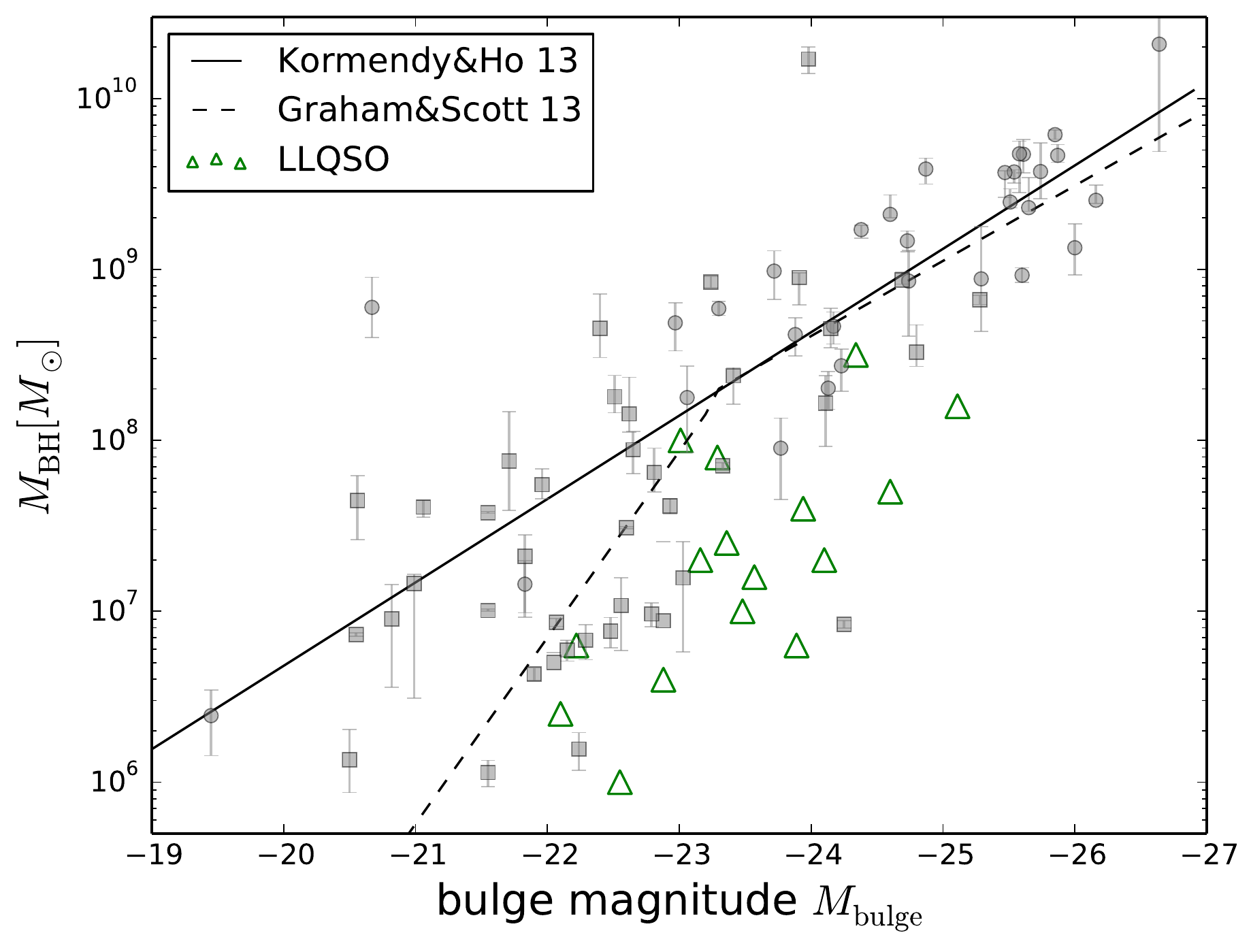}
\end{center}
\caption{The diagram shows a possible evolution scenario in the $M_\mathrm{BH}-L_\mathrm{bulge}$ diagram: Star formation sets in first and leads to an overluminosity (compared to inactive galaxies lying on the relation) of the bulges. Black hole growth will then shift the sources up. At the same time the starburst ages and the bulge luminosity will decrease.}
\label{fig:ML}
\end{figure}

A possible scenario to explain the offset of active galaxies from the $M_\mathrm{BH}-L_\mathrm{bulge}$ relation, illustrated in Fig.~\ref{fig:ML}, is the following:
\begin{enumerate}[label=(\roman*)]
\item Bulges grow first. Due to the young stellar populations, also the mass-to-light ratio is lower which results in a significant higher luminosity of the bulge. This shifts the galaxy to the right in the $M_\mathrm{BH}-L_\mathrm{bulge}$ relation.
\item Black hole growth is delayed by a few $100\,\mathrm{Myr}$ \citep{2007ApJ...671.1388D,2010MNRAS.405..933W}. In this phase, the galaxy appears as an active galaxy (Seyfert or QSO). The galaxy will be shifted up towards higher black hole masses in the $M_\mathrm{BH}-L_\mathrm{bulge}$ relation.
\item At the same time, the stellar population in the bulge will age. This means the mass-to-light ratio increases and the total bulge luminosity decreases. At the end of this evolution, the galaxy will end up with a slightly higher bulge luminosity (because the total stellar mass increased but the mass-to-light ratio increased as well) and a slightly higher black hole mass (due to mass accretion onto the central BH during the AGN phase) with lower Eddington ratio.
\end{enumerate}

In this scenario, active galaxies should always be located under the $M_\mathrm{BH}-L_\mathrm{bulge}$ relation. However the magnitude of the offset depends on the stage in the scenario. Indeed, studying a sample of $\sim 15$ low-luminosity QSOs \citep{2014A&A...561A.140B,2016A&A...587A.138B}, we find all sources systematically shifted below the relation (both, the linear and the near-quadratic one). This is consistent with most of the above mentioned previous studies, supporting the discussed scenario.

In a sample of AGN with BH masses from reverberation mapping, \cite{2009ApJ...694L.166B} note a slope of the $M_\mathrm{BH}-L_\mathrm{bulge}$ relation shallower than unity, however, they do not find a clear offset from the relation. We note that these objects are mostly at higher BH masses and might be in a more evolved phase, therefore closer to the relations. Furthermore, their study is based on a galfit-decomposition of optical HST images, while ours is based on BUDDA-decompositions of ground-based near-infrared images. This underlines the need for systematic studies that quantify possible technical biases originating from different fitting algorithms and wavelengths. Furthermore, we note that our scenario might only work for local AGN at the low BH mass end. The discussion of mergers or the redshift evolution of the scaling relations is beyond the scope of this article.

\section{Outlook: Insight from near-infrared integral-field spectroscopy}

\begin{figure}[t]
\begin{center}
\includegraphics[width=0.4\linewidth]{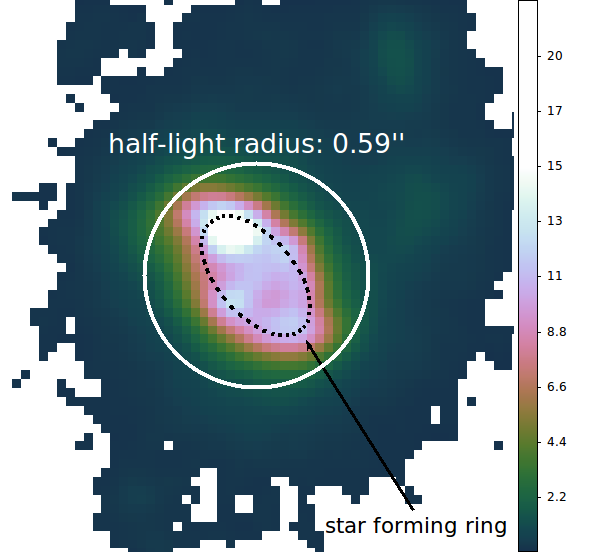}
\includegraphics[width=0.55\linewidth]{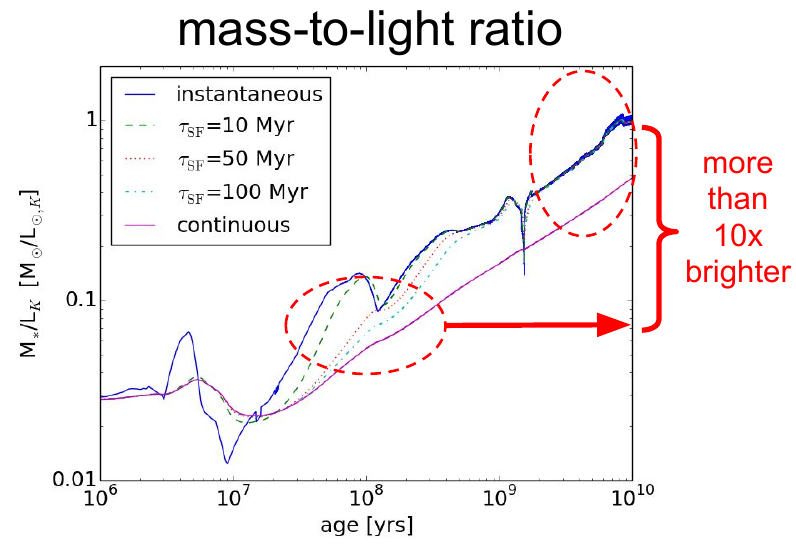}
\end{center}
\caption{\emph{Left:} Pa$\alpha$ flux distribution of the LLQSO HE 1029-1831. A circumnuclear ring with radius $0.3'' \approx 240\,\mathrm{pc}$ is visible. The ring contains younger stellar populations which come from a decaying starburst ($\tau=50-100\,\mathrm{Myr}$) that began $100-200\,\mathrm{Myr}$ before. \emph{Right:} Simulations with \textsc{Starburst99} indicate that the stellar populations in the ring can be up to ten times brighter than old stellar populations that are usually expected in bulges.}
\label{fig:he1029}
\end{figure}

An important way to test this scenario and understand the evolution of galaxies in the $M_\mathrm{BH}-L_\mathrm{bulge}$ relation is an accurate, spatially resolved determination of stellar populations and therefore mass-to-light ratios in the centres of active galaxies. Only in the near-infrared, integral-field spectrographs (such as SINFONI@ESO-VLT, OSIRIS@Keck or NIFS@GEMINI) equipped with adaptive optics systems can reach angular resolutions in the $\sim 100\,\mathrm{mas}$ range. 

In our pilot study with SINFONI, we study the central kiloparsec of the gas-rich low-luminosity QSO HE 1029-1831 ($z=0.0404$) at a resolution of $\sim 150\,\mathrm{mas}$ \citep{2015A&A...575A.128B}. Only with adaptive optics, we can resolve a circumnuclear star formation ring (and spatially disentangle it from the AGN emission) with radius $0.3''\approx 240\,\mathrm{pc}$. We compare observables such as emission from the hydrogen recombination line Br$\gamma$ and the forbidden transition [Fe\textsc{ii}] (which is a measure of the supernova rate) to simulations from \textsc{Starburst99} \citep{2014ApJS..212...14L} to estimate the age of the stellar populations in the ring. We find that the bulge is dominated by young stellar populations in the ring that originate from a decaying starburst with $\tau=50-100\,\mathrm{Myr}$ that began $100-200\,\mathrm{Myr}$ before (Fig.~\ref{fig:he1029}, left). Figure \ref{fig:he1029} (right) shows that these populations can be brighter by a factor of 10 compared to old, evolved stellar populations which are usually expected in bulges. The following rough estimation demonstrates that these populations could easily increase the bulge luminosity by one magnitude: From Fig.~\ref{fig:he1029} we see that the star forming ring fills $\sim 50\%$ of the half-light radius. By definition this again contains $50\%$ of the bulge luminosity. If $25\%$ of the bulge luminosity are increased by a factor of 10, this corresponds to a total bulge luminosity increase by a factor of $0.25\times 10+0.75\times 1= 3.25$ which is more than a magnitude difference. This is only a rough upper limit. For a more accurate calculation, for example, the flux distribution needs to be deconvolved. The fact that a starburst is significantly increasing the bulge luminosity and that it is already decaying, are consistent with the described scenario. 

In a next step, it will be important to better constrain the involved time-scales. In particular, the duration of the AGN cycle is essential to estimate the increase of the BH mass due to accretion. Furthermore, it has to be investigated in more detail how age estimates of the stellar populations are affected by underlying old bulge populations and which effect this has on the determination of the time-scales for the starburst and the probable time lag between starburst and AGN activity.

For this, larger and systematic studies of this kind are necessary. For example, an international team of researchers recently started the CARS survey (\emph{http://www.cars-survey.org/}) which aims to deliver a spatially resolved and multiwavelength legacy survey of 40 nearby quasars. With this survey, we will be able to study the stellar populations and star formation properties of bulges in active galaxies in greater detail.

\acknowledgments 
{I thank M\'onica Valencia-S. for a thorough reading of the manuscript and many suggestions and advice that helped to improve this paper. I further thank the two referees for their reports that helped me clarify several passages of the paper. Furthermore, I am thankful to our colleagues in Prague, especially Prof.~Vladim\'ir Karas, for their great hospitality. The stay of GB in Prague was supported by the collaboration programme between the University of Cologne and the Charles University in Prague and the Czech Academy of Sciences-DAAD exchange programme between the University of Cologne and the Astronomical Institute of the Academy of Sciences in Prague. This work was carried carried out within the Collaborative Research Centre 956, sub-project A2, funded by the Deutsche Forschungsgemeinschaft (DFG). GB is a member of the Bonn-Cologne Graduate School of Physics and Astronomy (BCGS).}

\bibliographystyle{egs}
\bibliography{wds}       

\end{article}
\end{document}